# Relativistic five-quark equations and the LHCb pentaquarks.

Gerasyuta S.M.[1] , Kochkin V.I.[2]

Department of Physics, St. Petersburg state Forest Technical University, Institutski Per. 5, St. Petersburg, 194021, Russia,

1. E-mail: gerasyuta@sg6488.spb.edu
2. E-mail: vik@efa.ru

## Abstract

The relativistic five-quark equations are found in the framework of the dispersion relation technique. The solutions of these equations using the method based on the extraction of the leading singularities of the amplitudes are obtained. The five-quark amplitudes for the hidden-charm pentaquarks including the u, d, c - quarks are calculated. The poles of these amplitudes determine the masses of $P_c$ and $P_c^*$ pentaquarks. The masses of pentaquarks with $J^P = \frac{3}{2}^-, \frac{5}{2}^+$ are calculated.

# I. Introduction.

Recent observation of the exotic baryon states with the hidden-charm by LHCb collaboration [1] has raised great interest in hadron physics. The states cannot by an ordinary three-quark baryon, and the minimal quark content is $uudc\bar{c}$.

The discovery of $P_c^+$, $P_c^{+*}$ has triggered extensive theoretical studies to understand the structure of these pentaquarks [2 - 15]. They are kinematic effects due to rescattering among different channels [2 - 5]. On the other hand, they could be bound states formed from open-charm baryon and charmed meson [6 - 10]. The exotic baryon with hidden charm as antiquark-diquark-diquark are considered [11, 12].

The $J^P$ assignment are not yet determined definitively, but the first Assignment is obtained with $P_c$(4380) and $P_c^*$(4450) as $\frac{3}{2}^-$ and $\frac{5}{2}^+$ states, respectively. The second Assignment is the $\frac{3}{2}^+$(4380) and the $\frac{5}{2}^-$(4450). Our calculation suggests the third Assignment the $\frac{5}{2}^+$(4380) and the $\frac{3}{2}^-$(4450).

In series of papers [16 – 20], a practical treatment of relativistic three-hadron systems has been developed. The physics of the three-hadron system is usefully described in terms of the pair wise interaction among the three particles. The theory is based on the two principles of unitarity and analyticity, applied to the two-body subenergy channels. The linear integral equations in a single variable are obtained for the isobar amplitudes. Instead of the quadrature, method of obtained the set of suitable functions are identified and used a basis for the expansion of desired solutions. By this means the coupled integral equations are solved in terms of simple algebra.

In our papers [21 – 24], relativistic generalization of three-body Faddeev equations was obtained in the form of dispersion relation in the pair energy of two interactions particles. The mass spectra of S-wave baryon including u, d, s, c - quarks were calculated by a method based on isolating the leading singularities in the amplitude. We searched for the approximate solution of integral three-quark equations by taking into account two-particle and triangle singularities, all the weaker ones being neglected. If we considered such an approximation, which corresponds to taking into account two-body and triangle singularities, and defined all the smooth functions in the middle point the physical region of the Dalitz plot, then the problem was reduced to solving a system of simple algebraic equations.

In the present paper, the relativistic five quark equation are found in the framework of coupled-channel formalism. We obtained the masses pentaquark $P_c$, $P_c^*$ in Assignment 3. Only this considered allows us to calculate the $P_c^*(4450)$ with $J^P = \frac{3}{2}^-$ as input. The other mass is $P_c(4430)$, $J^P = \frac{5}{2}^+$. The paper is organized as follows. After this introduction, we discussed the five-quark amplitudes, which contain two c – quarks, and three u, d quarks (Sec. 2). In Sec. 3 we report our numerical results (Table I).

## II. Five-quark amplitudes with hidden charm.

The five-quark equations in the framework of the dispersion relation technique are derived. Only planar diagrams are used; the other diagrams due to the rules of $1/N_c$ expansion are neglected [25 - 27]. The correct equations for the amplitude taking into account all possible subamplitudes. Then one should represent a five-particle amplitude as a sum of ten subamplitudes: $A = A_{12} + A_{13} + A_{14} + A_{15} + A_{23} + A_{24} + A_{25} + A_{34} + A_{35} + A_{45}$.

We can consider only one group of diagrams and the amplitude corresponding to them, for example $A_{12}$. The set of diagrams associated with the amplitude $A_{12}$ corresponds to amplitudes: $A_1(s,s_{1234},s_{12},s_{34})$, $A_2(s,s_{1234},s_{15},s_{34})$, $A_3(s,s_{1234},s_{23},s_{234})$, $A_4(s,s_{1234},s_{35},s_{345})$ (Fig. 1). For the sake of simplicity the $P_c(4430)$ with $J^P = \frac{5}{2}^+$ are shown. The coefficients are determined by the permutation of quarks [28, 29].

In order to represent the subamplitudes $A_1(s,s_{1234},s_{12},s_{34})$, $A_2(s,s_{1234},s_{15},s_{34})$, $A_3(s,s_{1234},s_{23},s_{234})$ and $A_4(s,s_{1234},s_{35},s_{345})$ in the form of a dispersion relation we define the amplitudes of quark-quark and quark-antiquark interaction $b_n(s_{ik})$. The pair quarks amplitudes $q\bar{q} \to q\bar{q}$ and $qq \to qq$ using the dispersion N/D method with the input four-fermion interaction with quantum numbers of the gluon [30] are considered. In our relativistic quark model [30] the pair quarks amplitude in have the follow form:

$$b_n(s_{ik}) = \frac{G_n^2(s_{ik})}{1 - B_n(s_{ik})}, \qquad (1)$$

$$B_n(s_{ik}) = \int_{(m_1+m_2)^2}^{\Lambda_n} \frac{ds'_{ik}}{\pi} \frac{\rho_n(s'_{ik}) G_n^2(s'_{ik})}{s'_{ik} - s_{ik}}. \qquad (2)$$

Here $s_{ik}$ is the two-particle subenergy squared, $s_{ijk}$ corresponds to the energy squared of particles $i$, $j$, $k$, $s_{ijkl}$ is the four-particle subenergy squared and $s$ is the system total energy squared. $G_n(s_{ik})$ are the quark-quark and quark-antiquark vertex functions (Table II). $B_n(s_{ik})$, $\rho_n(s_{ik})$ are the Chew-Mandelstam function with cut-off $\Lambda_n$ [31] and the phase space respectively:

$$\rho_n(s_{ik}, J^{PC}) = \left( \alpha(J^{PC},n) \frac{s_{ik}}{(m_i+m_k)^2} + \beta(J^{PC},n) + \delta(J^{PC},n) \frac{(m_i-m_k)^2}{s_{ik}} \right) \times$$

$$\times \frac{\sqrt{[s_{ik} - (m_i+m_k)^2][s_{ik} - (m_i-m_k)^2]}}{s_{ik}}$$

The coefficients $\alpha(J^{PC},n)$, $\beta(J^{PC},n)$ and $\delta(J^{PC},n)$ are given in Table III. Here n=1 corresponds to a $qq$-pair with $J^P = 0^+$ in the $\bar{3}_c$ color state, n=2 describes a $qq$-pair with $J^P = 1^+$ in the $\bar{3}_c$ color state and n=3 defines the $q\bar{q}$-pairs corresponding to mesons with quantum numbers: $J^{PC} = 1^{++}, 1^{--}$.

In the case in question the interacting quarks do not produce a bound state, therefore the integration in Eqs. (3) - (6) is carried out from the threshold $(m_i + m_k)^2$ to the cut-off $\Lambda_n$. The system of integral equations systems, corresponding to Fig. 1 (the meson state with $J^{PC} = 1^{++}$ and diquark with $J^P = 1^+$ and $J^P = 0^+$) can be described as:

$$A_1(s, s_{1234}, s_{12}, s_{34}) = \frac{\lambda_1 B_3(s_{12}) B_2(s_{34})}{[1 - B_3(s_{12})][1 - B_2(s_{34})]} + 6\hat{J}_2(3,2) A_3(s, s_{1234}, s'_{23}, s'_{234}) + \\ + 4\hat{J}_1(2) A_4(s, s_{1234}, s'_{35}, s'_{345}) \tag{3}$$

$$A_2(s, s_{1234}, s_{15}, s_{34}) = \frac{\lambda_2 B_3(s_{15}) B_2(s_{34})}{[1 - B_3(s_{15})][1 - B_2(s_{34})]} + 6\hat{J}_2(3,2) A_4(s, s_{1234}, s'_{35}, s'_{345}) + \\ + 4\hat{J}_1(2) A_3(s, s_{1234}, s'_{23}, s'_{234}) \tag{4}$$

$$A_3(s, s_{1234}, s_{23}, s_{234}) = \frac{\lambda_3 B_1(s_{23})}{1 - B_1(s_{23})} + 2\hat{J}_3(1) A_1(s, s_{1234}, s'_{12}, s'_{34}), \tag{5}$$

$$A_4(s, s_{1234}, s_{35}, s_{345}) = \frac{\lambda_4 B_1(s_{35})}{1 - B_1(s_{35})} + \hat{J}_3(1) A_2(s, s_{1234}, s'_{15}, s'_{34}), \tag{6}$$

were $\lambda_i$ are the current constants. We introduce the integral operators:

$$\hat{J}_1(l) = \frac{G_l(s_{12})}{[1 - B_l(s_{12})]} \int_{(m_1+m_2)^2}^{\Lambda_l} \frac{ds'_{12}}{\pi} \frac{G_l(s'_{12})\rho_l(s'_{12})}{s'_{12} - s_{12}} \int_{-1}^{+1} \frac{dz_1}{2}, \tag{7}$$

$$\hat{J}_2(l, p) = \frac{G_l(s_{12}) G_p(s_{34})}{[1 - B_l(s_{12})][1 - B_p(s_{34})]} \times \\ \times \int_{(m_1+m_2)^2}^{\Lambda_l} \frac{ds'_{12}}{\pi} \frac{G_l(s'_{12})\rho_l(s'_{12})}{s'_{12} - s_{12}} \int_{(m_3+m_4)^2}^{\Lambda_p} \frac{ds'_{34}}{\pi} \frac{G_p(s'_{34})\rho_p(s'_{34})}{s'_{34} - s_{34}} \int_{-1}^{+1} \frac{dz_3}{2} \int_{-1}^{+1} \frac{dz_4}{2}, \tag{8}$$

$$\hat{J}_3(l) = \frac{G_l(s_{12}, \tilde{\Lambda})}{1 - B_l(s_{12}, \tilde{\Lambda})} \times \\ \times \frac{1}{4\pi} \int_{(m_1+m_2)^2}^{\tilde{\Lambda}} \frac{ds'_{12}}{\pi} \frac{G_l(s'_{12}, \tilde{\Lambda})\rho_l(s'_{12})}{s'_{12} - s_{12}} \int_{-1}^{+1} \frac{dz_1}{2} \int_{-1}^{+1} dz \int_{z_2^-}^{z_2^+} dz_2 \frac{1}{\sqrt{1 - z^2 - z_1^2 - z_2^2 + 2zz_1 z_2}}, \tag{9}$$

Where $l$, $p$ are equal 1 - 3. Here $m_i$ is a quark mass.

In the Eqs. (7) and (9) $z_1$ is the cosine of the angle between the relative momentum of the particles 1 and 2 in the intermediate state and the momentum of the particle 3 in the final state, taken in the c.m. of particles 1 and 2. In the Eq. (9) $z$ is the cosine of the angle between the momenta of the particles 3 and 4 in the final state, taken in the c.m. of particles 1 and 2. $z_2$ is the cosine of the angle between the relative momentum of particles 1 and 2 in the intermediate state and the momentum of the particle 4 in the final state, is taken in the c.m. of particles 1 and 2. In the Eq. (8): $z_3$ is the cosine of the angle between relative momentum of particles 1 and 2 in the intermediate state and the relative momentum of particles 3 and 4 in the intermediate state, taken in the c.m. of particles 1 and 2. $z_4$ is the cosine of the angle between the relative momentum of the particles 3 and 4 in the intermediate state and that of the momentum of the particle 1 in the intermediate state, taken in the c.m. of particles 3, 4.

We can pass from the integration over the cosines of the angles to the integration over the subenergies [32].

Let us extract two-particle singularities in the amplitudes $A_1(s,s_{1234},s_{12},s_{34})$, $A_2(s,s_{1234},s_{15},s_{34})$, $A_3(s,s_{1234},s_{23},s_{234})$ and $A_4(s,s_{1234},s_{35},s_{345})$:

$$A_1(s,s_{1234},s_{12},s_{34}) = \frac{\alpha_1(s,s_{1234},s_{12},s_{34})B_3(s_{12})B_2(s_{34})}{[1-B_3(s_{12})][1-B_2(s_{34})]}, \tag{10}$$

$$A_2(s,s_{1234},s_{15},s_{34}) = \frac{\alpha_2(s,s_{1234},s_{15},s_{34})B_3(s_{15})B_2(s_{34})}{[1-B_3(s_{15})][1-B_2(s_{34})]}, \tag{11}$$

$$A_3(s,s_{1234},s_{23},s_{234}) = \frac{\alpha_3(s,s_{1234},s_{23},s_{234})B_1(s_{23})}{1-B_1(s_{23})}, \tag{12}$$

$$A_4(s,s_{1234},s_{35},s_{345}) = \frac{\alpha_4(s,s_{1234},s_{35},s_{345})B_1(s_{35})}{1-B_1(s_{35})}, \tag{13}$$

We do not extract three- and four-particle singularities, because they are weaker than two-particle singularities.

The functions $\alpha_1(s,s_{1234},s_{12},s_{34})$, $\alpha_2(s,s_{1234},s_{15},s_{34})$, $\alpha_3(s,s_{1234},s_{23},s_{234})$ and $\alpha_4(s,s_{1234},s_{35},s_{345})$ are smooth functions of $s_{ik}$, $s_{ijk}$, $s_{ijkl}$, $s$ as compared with the singular part of the amplitudes, hence they can be expanded in a series in the singularity point and only the first term of this series should be employed further. Using this classification one define the reduced amplitudes $\alpha_1$, $\alpha_2$, $\alpha_3$, $\alpha_4$ as well as the B-functions in the middle point of the physical region of Dalitz-plot at the point $s_0$:

$$s_0^{ik} = s_0 = \frac{s + 3\sum_{i=1}^{5} m_i^2}{0.25 \sum_{\substack{i,k=1 \\ i \neq k}}^{5} (m_i + m_k)^2} \quad (14)$$

$$s_{123} = 0.25 s_0 \sum_{\substack{i,k=1 \\ i \neq k}}^{3} (m_i + m_k)^2 - \sum_{i=1}^{3} m_i^2, \quad s_{1234} = 0.25 s_0 \sum_{\substack{i,k=1 \\ i \neq k}}^{4} (m_i + m_k)^2 - 2\sum_{i=1}^{4} m_i^2$$

Such a choice of point $s_0$ allows one to replace the integral Eqs. (3) - (6) (Fig. 1) by the algebraic equations (15) - (18) respectively:

$$\alpha_1 = \lambda_1 + 6\hat{J}_2(3,2,1)\alpha_3 + 4\hat{J}_1(2,1)\alpha_4, \quad (15)$$

$$\alpha_2 = \lambda_2 + 6\hat{J}_2(3,2,1)\alpha_4 + 4\hat{J}_1(2,1)\alpha_3, \quad (16)$$

$$\alpha_3 = \lambda_3 + 2\hat{J}_3(1,2,3)\alpha_1, \quad (17)$$

$$\alpha_4 = \lambda_4 + \hat{J}_3(1,2,3)\alpha_2, \quad (18)$$

We use the functions $J_1(l,p)$, $J_2(l,p,r)$, $J_3(l,p,r)$ ($l,p,r = 1, 2, 3$):

$$J_1(l,p) = \frac{G_l^2(s_0^{12})B_p(s_0^{13})}{B_l(s_0^{12})} \int_{(m_1+m_2)^2}^{\Lambda_l} \frac{ds'_{12}}{\pi} \frac{\rho_l(s'_{12})}{s'_{12} - s_0^{12}} \int_{-1}^{+1} \frac{dz_1}{2} \frac{1}{1 - B_p(s'_{13})}, \quad (19)$$

$$J_2(l,p,r) = \frac{G_l^2(s_0^{12})G_p^2(s_0^{34})B_r(s_0^{13})}{B_l(s_0^{12})B_p(s_0^{34})} \times$$

$$\times \int_{(m_1+m_2)^2}^{\Lambda_l} \frac{ds'_{12}}{\pi} \frac{\rho_l(s'_{12})}{s'_{12} - s_0^{12}} \int_{(m_3+m_4)^2}^{\Lambda_p} \frac{ds'_{34}}{\pi} \frac{\rho_p(s'_{34})}{s'_{34} - s_0^{34}} \int_{-1}^{+1} \frac{dz_3}{2} \int_{-1}^{+1} \frac{dz_4}{2} \frac{1}{1 - B_r(s'_{13})} \quad (20)$$

$$J_3(l,p,r) = \frac{G_l^2(s_0^{12}, \tilde{\Lambda})B_p(s_0^{13})B_r(s_0^{24})}{1 - B_l(s_0^{12}, \tilde{\Lambda})} \frac{1 - B_l(s_0^{12})}{B_l(s_0^{12})} \times$$

$$\times \frac{1}{4\pi} \int_{(m_1+m_2)^2}^{\tilde{\Lambda}} \frac{ds'_{12}}{\pi} \frac{\rho_l(s'_{12})}{s'_{12} - s_0^{12}} \int_{-1}^{+1} \frac{dz_1}{2} \int_{-1}^{+1} dz \int_{z_2^-}^{z_2^+} dz_2 \frac{1}{\sqrt{1 - z^2 - z_1^2 - z_2^2 + 2zz_1z_2}} \frac{1}{[1 - B_p(s'_{13})][1 - B_r(s'_{24})]} \quad (21)$$

The other choices of point $s_0$ do not change essentially the contributions of $\alpha_1$, $\alpha_2$, $\alpha_3$ and $\alpha_4$ therefore we omit the indexes $s_0^{ik}$. Since the vertex functions depend only slightly on energy it is possible to treat them as constants in our approximation. The integration contours of function $J_1, J_2, J_3$ are given in paper [32].

The solutions of the system of equations are considered as:

$$\alpha_i(s) = F_i(s, \lambda_i) / D(s), \qquad (22)$$

where zeros of $D(s)$ determinants define the masses of bound states of pentaquark baryons. $F_i(s, \lambda_i)$ are the functions of $s$ and $\lambda_i$. The functions $F_i(s, \lambda_i)$ determine the contributions of subamplitudes to the pentaquark baryon amplitude.

### III. Calculation results.

The poles of the reduced amplitudes $\alpha_1$, $\alpha_2$, $\alpha_3$, $\alpha_4$ correspond to the bound states and determine the masses of pentaquarks with hidden charm (Assignment 3). We used the graphic representation of the equation $J^P = \frac{5}{2}^+$ for the sale a simplicity (Fig. 1). But as fit we considered is equations for the $J^P = \frac{3}{2}^-$. The experimental mass value of $P_c^* = 4450$ for the states with $I = \frac{1}{2}$ and $J^P = \frac{3}{2}^-$ is calculated. Our model uses only three parameters. The cutoff $\Lambda = 10$ is taken for the previous paper. The gluon constant is determined by the massive state $J^P = \frac{3}{2}^-$. The quark masses are $m_{u,d} = 385+25$ MeV, $m_c = 1586+25$ MeV. The shift $\Delta = 25$ MeV take into account the confiment. We calculated the mass the second state $J^P = \frac{5}{2}^+$. This mass is equal to 4430, this state have the large width and similar to state 4380 MeV. We considered the $J^P = \frac{5}{2}^+$ (4430) and $J^P = \frac{3}{2}^-$ (4450). The P - parity state is smaller than S-wave.

IV. Conclusions.

The Assignment 3 allows as considered two masses of pentaquark with $J^P = \frac{5}{2}^+$ and $J^P = \frac{3}{2}^-$. The second state described the mass $J^P = \frac{5}{2}^+$, which is different from experimental bat $J^P = \frac{5}{2}^+$ with the mass 4430. Experimental date M=4380 and the width not small. The positive parity state are smaller the $J^P = \frac{3}{2}^-$.

Acknowledgments.

The reported study was supported by RFBR, research project № 13-02-91154.

Table I. Pentaquark masses with isospin $I = \frac{1}{2}$

| $J^P$ | Mass, MeV |
|---|---|
| $\frac{5}{2}^+$ | 4430 |
| $\frac{3}{2}^-$ | 4450 |

Parameters of model: quark mass $m_{u,d} = 410$ MeV, $m_c = 1611$ MeV, $\Delta = 25$ MeV; cut-off parameter $\Lambda = 10$; gluon coupling constant $g = 0.977$.

Table II. Vertex functions

| $J^{PC}$ | $G_n^2$ |
|---|---|
| $0^+$ (n=1) | $4g/3 - 2g(m_i + m_k)^2/(3s_{ik})$ |
| $1^+$ (n=2) | $2g/3$ |
| $1^{++}$ (n=3) | $4g/3$ |
| $1^{--}$ (n=3) | $4g/3$ |

Table III. Coefficient of Chew-Mandelstam functions for n = 3 (meson states) and diquarks n = 1 ($J^P = 0^+$), n = 2 ($J^P = 1^+$).

| $J^{PC}$ | n | $\alpha(J^{PC},n)$ | $\beta(J^{PC},n)$ | $\delta(J^{PC},n)$ |
|---|---|---|---|---|
| $0^+$ | 1 | 1/2 | -e/2 | 0 |
| $1^+$ | 2 | 1/3 | 1/6-e/3 | -1/6 |
| $1^{++}$ | 3 | 1/2 | -e/2 | 0 |
| $1^{--}$ | 3 | 1/3 | 1/6-e/3 | -1/6 |

$e = (m_i - m_k)^2 / (m_i + m_k)^2$

Figure captions

Fig.1. Graphic representation of the equations for the five-quark subamplitudes $A_k$ ($k = 1$-4).

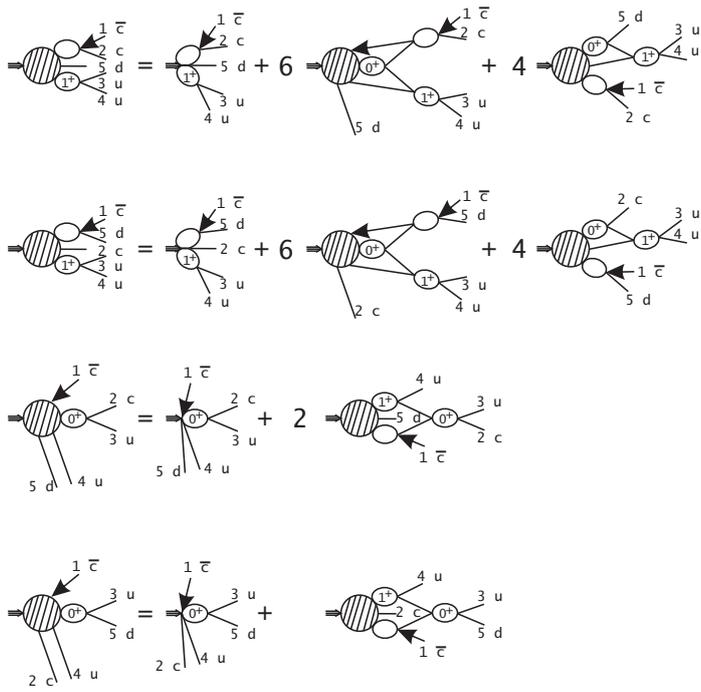

Fig.1